\documentclass{llncs}

\usepackage[latin1]{inputenc}
\usepackage{graphicx}
\usepackage{color}
\usepackage{amssymb}
\usepackage{amsmath}
\catcode`"=\active
\def"#1"{``#1''}
\catcode`*=\active
\def*#1*{\textbf{#1}}

\begin{document}

\title{The influence of risk perception in epidemics: 
a cellular agent model}

\author{
  Luca Sguanci\thanks{luca.sguanci@unifi.it}\inst{1}\and
  Pietro Li\`o\thanks{pietro.lio@cl.cam.ac.uk}\inst{2}\and
  Franco Bagnoli\thanks{franco.bagnoli@unifi.it}\inst{1}
}
\institute{
  Dept. Energy, Univ. of Florence, Via
  S. Marta 3, 50139 Firenze, Italy; \\
  also CSDC and INFN, sez. Firenze.   
 \and
  Computer Laboratory, University of Cambridge, CB3 0FD
  Cambridge, UK.
}

\maketitle

\begin{abstract}
Our work stems from the consideration that the spreading of a
disease is modulated by the individual's perception of the
infected neighborhood and his/her strategy to avoid being infected
as well. 
We introduced a general ``cellular agent'' model that accounts for a hetereogeneous and variable network of connections.
The probability of infection is assumed to depend on the perception that an individual has about the spreading of the disease in her local neighborhood and on broadcasting media. In the one-dimensional homogeneous case the model reduces to the DK one, while for long-range coupling the dynamics exhibits large fluctuations that may lead to the  complete extinction of the disease.
\end{abstract}


\section{Introduction}
\label{sec:Introduction}

In "Les rois thaumaturges: étude sur le caractère surnaturel attribué à la
puissance royale particulièrement en France et en Angleterre" the historian
Marc Bloch ~\cite{Bloch} wrote that until about 1700, sick people in England
and France tried to be touched by the king who they believed was a
miraculous physician whose mere touch would cure physical illness.
Since then, much time has passed, we do no more touch the king but we
still have to face with illness and different pathologies. Now that we
know viruses and bacteria, we are addressing the issue of studying the
mutual influences between collective behaviour, disease spreading and
viral evolution. In fact, HIV epidemics has changed many of our sexual
and social behaviors~\cite{LA2003} and selection on viral strains has been in act by social groups~\cite{Schliekelman01,Novembre05}. Zanotto and
collaborators~\cite{ZH1996} have shown that viral evolution depends on
differences in modes of dispersal, propagation, and changes in the size
of host populations. They also suggest a link between the growing and
fluidity of the human population and its exposure to an expanding range
of increasingly diverse viral strains.

Understanding the role of social behaviour has potentiality of giving
better answers to the pressing public health questions about whether
and how we can contain or slow the spread of an emerging epidemics to give time for vaccine development.  Moreover, the
understanding of key properties of contact networks may allow to reduce
disease transmission, avoid both costly and time consuming universal
vaccination or leaving hidden pockets of poor coverage that will seed
again the epidemics.

Previous epidemiological models have investigated the effect of a wide
variety of parameters, such as use of antiviral agents, super
spreaders and individual variation~\cite{LG2005}, quarantine and
pre-vaccination to contain the spread of disease at source.  However,
an outmost important factor that has been ignored so far is how the
perception of the epidemics, as perceived from a neighborhood (short
range information contacts) or from the media (long range), will
change the diffusion parameters.

Here we concentrate on the study of the risk perception on disease
spreading in the case of a homogeneous population. Although
spatial variables can play a major role, it is important to study
average statistical properties (mean field analysis) before taking
into consideration more complex geometries. In general,
populations do not experience full-mixing condition. However,
well-stirred conditions are recovered whenever conditions of
people crowdedness are considered or if it is possible to focus on
a given scale of observation. Noteworthy the former conditions
occur very frequently in urban contexts, for example in tubes and
buses at peak times and aerial spreading of cold-related virus
particles from coughing and sneezing disregards the casual
contact. Other examples are children in a nursery who have large
number of contacts during the day. On the other hand, we can
concentrate on a homogeneous scale of observation if we study
disease spreading in the hubs constituted by airports and train
stations. Similarly, if we are interested in the interplay between
cities and the countryside in disease evolution, we may address
the problem considering the interaction between those two distinct
entities, each characterized by homogeneous properties.

Different models for spreading of epidemics have been proposed, either
considering homogeneous populations~\cite{Murray,May}, or in the
framework of complex networks~\cite{Vespignani}. This kind of approach allows
assessing the relative importance of local and long-range contacts not only in
spreading the infection but also in spreading information on the infection
risk and thus potentially stand as a very useful tool for public health
managing and decision making processes.

The paper is organized as follows. In the next section we present
a general \emph{cellular agent}
model for the study of the perceptive dynamics of a
disease spreading. In section III we present the mean field
approximation of the model, then we present the results of the
performed simulations and finally we draw our conclusions.

\section{The Model: partying with your neighbors or stay home, 
spy them and read the news?}

We shall develop a quite general agent-based model, allowing age
classes (progression of the illness) and different types of
communication networks. We propose to use the term \emph{cellular
agent} for it, since it reduces to cellular automata for a regular
lattice of connections, but connections may also change in time. 

The single agent $i$ (representing an individual or a group of strongly
connected individuals like a family) is implemented as a set of (directional)
incoming connections, an internal state and an output state. Let us denote as
$M_{ij}$ the connection from site $j$ to site $i$. In our model a connection
represents the propensity of being infected, which is proportional to the
fraction of time spent together  by the two individuals $i$ and $j$, but also
depends on the type of contact. For this last reason, the connection needs not
to be symmetric: while it may be true for friendly contacts, the risk of being
infected is quite asymmetrical for professionals (nurses, physicians, etc.)
and also for parents vs.\ children, and so on.   In the simplest case of
unweighted connection, $M_{ij}\in\{0,1\}$, $k_i=\sum_j M_{ij}$ is the number
of neighbors and $s_i = \sum_j M_{ij} [\sigma_j \neq 0]$\footnote{We use the notation $[\text{\emph{statement}}]$ to indicate
the truth function, which gives 1 if \emph{statement} is true and 0
otherwise.} is the number of
infected neighbors. In the case of weighted connections, $s$ and $k$ are no
more integers. 
The network of connections may be fixed, or evolving in time. The degree (or connectivity) of a node is defined as the number of the incoming/outcoming links, while the degree distributions of a network, $P(k)$, represents the fraction of nodes with degree $k$. Many social networks have a scale-free structure~\cite{SocialNetworks}, and
this kind of networks can only be \emph{grown} using a connection rule. So, it
is natural to assume that new connections may be established, and old one
removed, following a dynamical rule. Actually, one could work with a
fully-connected network, and implement the evolution of connection as a rule
for the intensities $M_{ij}$ (possibly introducing a threshold value for the
efficacy of a connection), but this would be quite expensive in computer
terms. We limit the present investigations to fixed connection all of the same
intensity. 

We represent the internal state (progression of illness)  of the individual $i$ as a
bitstring $\sigma_i$.  Each bit in $\sigma$ (represented as a base-2 number) indicates
the presence of a given strain.  In this  way we can account for the geographic
distribution of different strains (important for immunization strategies), multiple
infections (co-infection or delayed re-infection) and recombination among strains. To
each possible value of $\sigma$ is associated an infection probability (infectivity)
$\tau(\sigma)$, with $\tau(0)=0$. The internal state contains also a time counter, for
timing the progression of the illness. In the present model, we simply assume that the
individual becomes healthy after a certain interval from the last infection. We do not
consider here immunization, nor the internal dynamics between infective pathogens and
the immune system~\cite{ImmuneSystemDynamics}.

The output state indicate if an individual is infective, and if it is
visibly ill. In this way we can represent incubation periods. In this
first study, we assume that the illness become visible the unit of
time (day) after infection, thus obtaining a parallel evolution. 

We assume that the probability of infection is proportional to the frequency of contacts
$M_{ij}$,  but that it is also modulated by the individual's \emph{perception} of the
percentage of infected people in her neighborhood as well as by the strategy for
avoiding being infected. If an individual realizes that a large
fraction of her neighbors is infected, or is alerted by broadcasting media, then she
may change her habits.  She may rise the level of precautions (thus lowering the effective infectivity of the
illness) or alter her connection patterns.  Since this last choice implies a large
rearrangement of individual lifestyle, this dramatic change is assumed to take place
only in extreme cases. However, even without changing lifestyle it is possible to lower
the infection probability by simply taking elementary precautions. We assume this to be
the most common reaction. Therefore we keep $M_{ij}$ constant during the simulation, but
make the infection probability of a single contact to vary according with the fraction
of infected people among the neighbors (weighted with the connection strengths) and with
the influence by media information.
We also assume that the recovering is immediate, and that the individual
becomes immediately susceptible. 

The perception (information) about the disease is written as $I(s,k)=\exp[-(H+Js/k)]$.
The parameter $J$ modulates individual's response to the the local infection load. The
role of the intensity of the external fields, like public healths alerts and media
influences, is accounted for by the $H$ parameter. In the following we assume $H=0$, but it's worth noting that this
parameter can play a major role in scenarios of low perception of the risk of infection.
This could be the case of infections  characterized by a long-asymptomatic phase, in
which many contacts occurs without the perception of any risk of being infected. In
such scenarios, $H$ turns out to be the only mean to downregulate the spreading of the
disease. 

The microscopic infection process is the following: for all the contacts of the  individual $i$, the bitstring $\sigma_i$  is OR-ed with
$\sigma_j$, the bitstring representing the neighboring individual $j$, if the
contact is effective in propagating the infection. This happens with a
probability $M_{ij} I(s_i, k_i)\tau(\sigma_j)$.

The total infection probability $p(s,k)$ is therefore
\begin{equation}
p_i= 1-\prod_j\left[1-M_{ij}I(s_i,k_i)\tau(\sigma_j)\right]. 
\label{eq:p}
\end{equation}

In the unweighted case, with single-valued connectivity, $P(k')=\delta_{k,k'}$, and assuming the same infectivity $\tau$ for all
strains, equation~\eqref{eq:p} becomes:
\begin{equation}
p_i=1-[1-k I(s_i, k_i)\tau]^{s_i}. 
\label{eq:I-simple}
\end{equation}

\section{Results}
\subsection{One Dimensional Case}
Here we consider the simplest case where $M_{ij}$ defines a 1D regular lattice with $k=2$ (nearest neighbors), and where all the contacts have the same strength. The status of node $i$ is represented by a single bit,  $\sigma_i=\{0,1\}$ and the
infectivity parameter, $\tau$, is single valued. 

This case can be mapped on the Domany-Kinzel model~\cite{DK}. This latter is defined as a one-dimensional totalistic cellular automaton with $k=2$, and its evolution rule depends on two parameters: $p_1$, the probability becoming infected if only one of the neighbors is infected, and $p_2$, the probability of being infected if both neighbors are infected. The correspondence with our model is therefore $p_1 = p(1,2)$ and $p_2=p(2,2)$.

\begin{figure}
\centering
\includegraphics[width=0.45\textwidth]{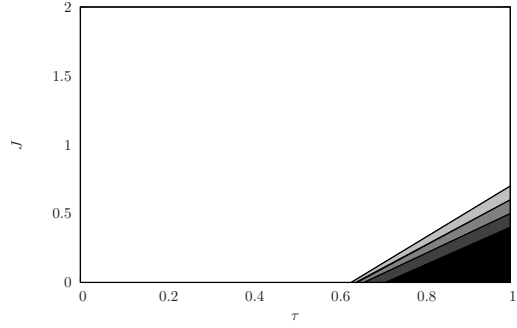}
\caption{Percentage of asymptotic infected population for the one-dimensional, $k=2$ case (1000 sites). White: no individual is infected, black: all individuals are infected.\label{fig:tauJ}}
\end{figure}

We have obtained the phase space ($H=0$) using the ($\tau$,$J$) parameters. The results are shown in Fig.~\ref{fig:tauJ}. The model exhibits a continuous transition (second-order) from a healthy state to the complete infection, as the infectivity increases. As far as $J$ is subsequently increased over a threshold value, the infection can no longer subsist and the population recovers completely from the disease.

\subsection{Long-range Case}

\subsubsection{Mean-field Approximation.}

The average asymptotic behavior of networks can be investigated by
means of mean field approach. Let us
call $N_k=N P(k)$ number of nodes with connectivity $k$; $\Omega_{k,k'}$, probability of a node 
with connectivity $k$ being connected to a node with connectivity
$k'$; $N_k c_k$: number of nodes with connectivity $k$ being
infected; $m=\sum_{k'} k' P(k')$ is the  average frequency of contact between two individuals.
If only one infective strain is considered, i.e.
$\tau(\sigma_j)=\tau$, the probability of being infected of a node
with connectivity $k$, $c_k$, is given by:

\begin{eqnarray}
c'_k & = &\sum_{s=1}^{k} \binom{k}{s}
  \left(
    \sum_{k_1^{'},k_2^{'},\ldots,k_s^{'}}
    (\Omega_{kk_1^{'}} c_{k_1^{'}}) \ldots (\Omega_{kk_s^{'}} c_{k_s^{'}})
  \right) \times \nonumber\\
  & & \left(
      \sum_{k_{s+1}^{'},\ldots,k_{k-s}^{'}} (\Omega_{k k_{s+1}^{'}} (1-c_{k_{s+1}^{'}}))
      \ldots (\Omega_{kk_{k-s}^{'}} (1-c_{k_{k-s}^{'}}))
    \right)
  \left[1-\left(1-m\ I(s,k) \tau \right)^s\right]=
  \nonumber\\
  &  & \sum_{s=1}^{k} \binom{k}{s} \left(\sum_{k'}\Omega_{kk'}c_{k'}\right)^s
    \left(\sum_{k'}\Omega_{kk'}(1-c_{k'})\right)^{k-s}
    \left[1-\left(1-m\ I(s,k) \tau \right)^s\right]
  \label{eq:meanfield}
\end{eqnarray}
If a non assortative network is considered, i.e.\ 
$\Omega_{k,k'}=N_{k'}/N=P(k')$:
\begin{equation}
  c'_k = \sum_{s=1}^{k} \binom{k}{s}
    \left(\sum_{k'}P(k')c_{k'}\right)^s
    \left(\sum_{k'}P(k')(1-c_{k'})\right)^{k-s}
    \left[1-\left(1-m\ I(s,k) \tau \right)^s\right]
\end{equation}
and, if $k$ is fixed, i.e. $P(k')=\delta_{k',k}$ and $m=k$,
\begin{equation}
c' =\sum_{s=1}^{k} \binom{k}{s} c^s (1-c)^{k-s} \left[1-\left(1-k\ I(s,k) \tau \right)^s\right]
\end{equation}

\subsubsection{Estimation of the infection reproductive rate.} 

A meaningful epidemiological parameter
is the basic reproductive rate, $R_{0}$, which is defined as the
mean number of infections caused by an infected individual in a
susceptible population~\cite{LH2005,LG2005}. This parameter
can be considered an epidemiological threshold. When $R_{0}<1$ , each
person who contracts the disease will infect fewer than one person
before dying or recovering, so the outbreak will cease. When
$R_{0}>1$, each person who gets the disease will infect more than one
person, so the epidemic will spread.

A more careful investigation of this parameter can lead to a
better insight in the dynamics of the epidemics, at the same time
allowing to assess the efficacy of different strategies of containment
on the spreading of the disease. For example Lloyd-Smith and
colleagues have shown that the distribution of individual
infectiousness around $R_{0}$ is often highly skewed~\cite{LG2005}.
Longini and collaborators have investigated bird flu pandemia
scenarios. They found that if $R_{0}$ was below 1.60, a prepared
response with targeted antivirals would have a high probability of
containing the disease. If pre-vaccination occurred, then targeted
antiviral prophylaxis could be effective for containing strains with
an $R_{0}$ as high as 2.1. Combinations of targeted antiviral
prophylaxis, pre-vaccination, and quarantine could contain strains
with an R0 as high as 2.4 ~\cite{LH2005}.

With reference to the model we propose, we can derive the expression of the basic reproductive ratio, by considering the variation of $c'$ with respect to $c$, when a small fraction of infected population is considered, i.e.
\begin{equation}
R_0=\lim_{c \to 0} \frac{\partial c'}{\partial c}  = k [p(1,k)\tau]
\end{equation}
In this way we recover the expression of the basic reproductive ratio, when a unitary mean time of infectivity per individual is considered. 
From this we derive the critical value of J, below which the fraction of infected individuals is different from zero,
\begin{equation}
J_{c}=k \log(k \tau)
\end{equation}

\subsubsection{Numerical Simulations.}

\begin{figure}
\centering
\includegraphics[width=0.45\textwidth]{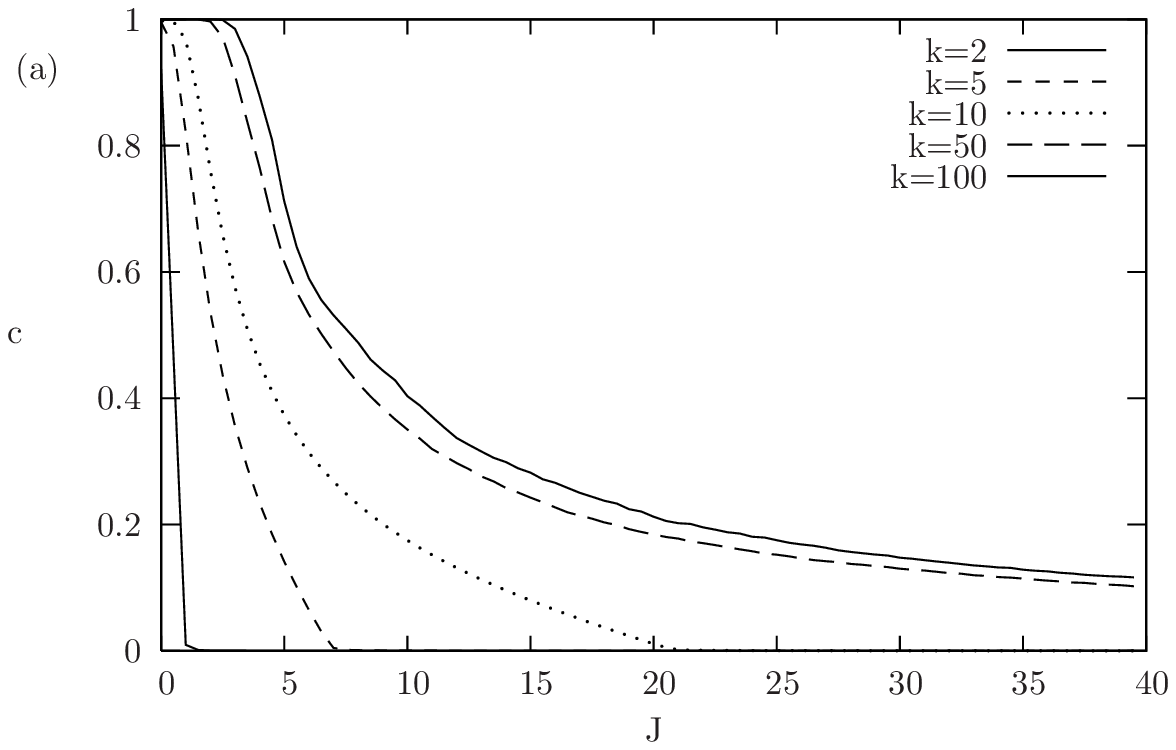}
\includegraphics[width=0.45\textwidth]{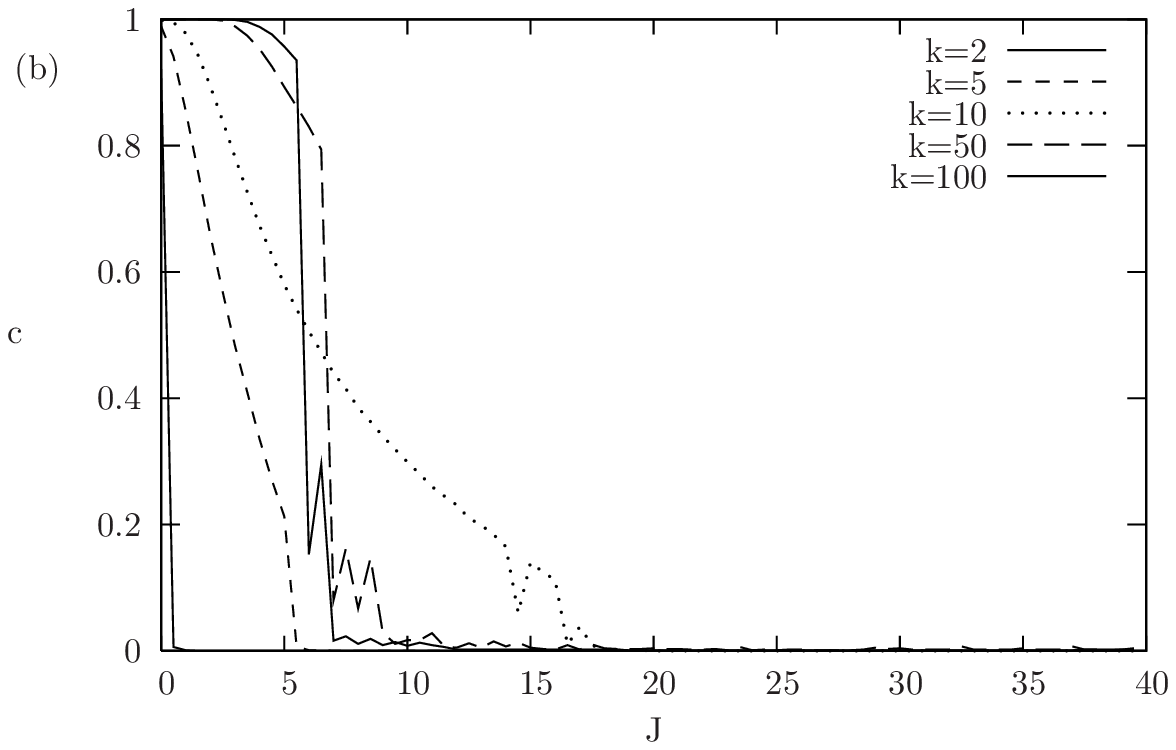}
\caption{Mean field computation (a) and numerical simulation (b, 100 sites) of the asymptotic fraction of infected sites, $c$.\label{fig:cJ}}
\end{figure}

\
To better characterize the role of the mean connectivity of individuals $k$ (randomly chosen), we plot the value of the fraction of infected individuals, $c$, as a function of $J$, for different values of $k$. 
In Fig.\ref{fig:cJ} we report the results of the numerical simulations for the mean field approximation of the model, plot (a), and for the microscopic dynamics, plot(b).
We can first notice that, for a growing number of neighbors, the fraction of infected individuals increases. This suggest that if we consider bounded the strength of the individual perception of the disease, an ever growing influence of the external field is necessary to keep low the number of infected individuals.
By comparing the mean field model with the microscopic dynamics a good agreement is shown for small values of $J$ (depending on $k$. For larger values of $J$, the infected population exhibits large coherent oscillations, that may may lead to a complete recover from the infection and to the disappearing of the epidemics.
In the mean field approximation, for increasing values of $k$, the model begins to show a high variation in the fraction of infected individuals, without reaching extinction. 

\begin{figure}
\centering
\includegraphics[width=0.45\textwidth]{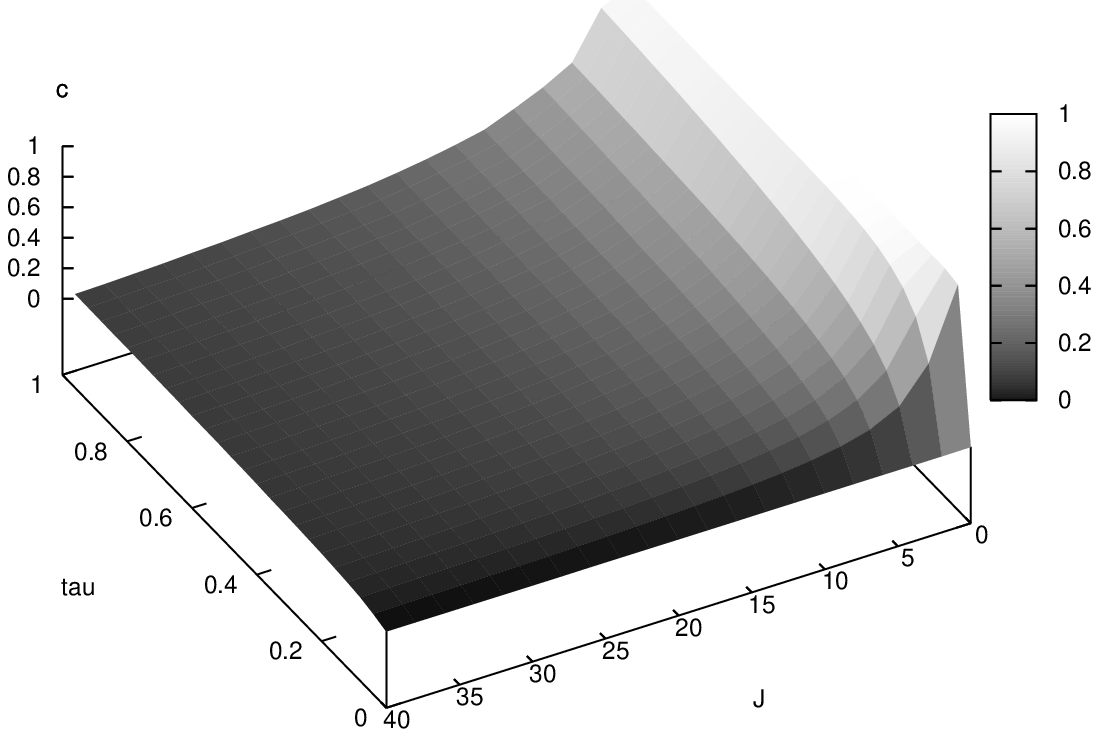}
\includegraphics[width=0.45\textwidth]{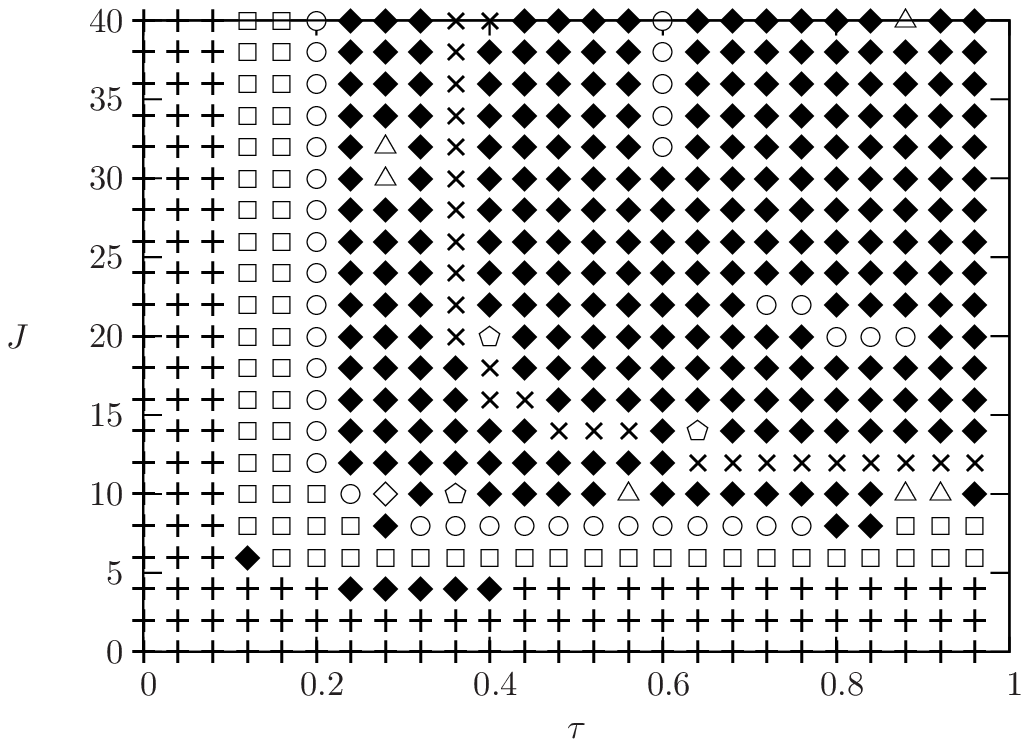}
\caption{Mean field asymptotic value of the fraction of infected sites (left) and  bifurcation diagram for $k=50$, $N=100$. Pluses: fixed points, empty squares: period-2, crosses: period-3,  empty circles: period-4, empty triangles: period-5, empty pentagons: period-6, empty diamonds: period-8, filled diamonds: chaotic orbits. \label{fig:cm-tauJ}}
\end{figure}

By keeping fixed the value of the mean connectivity and setting $H=0$, we analyzed the mean-field  phase space. In Fig.\ref{fig:cm-tauJ}, the case for $k=100$ is reported. It's interesting to observe the  rich dynamics exhibited by the model phase space, with oscillations and chaotic behavior.

\section{Inching Towards Reality}

As the amount of epidemiological data in the public domain grows,
so does the range of social science topics that it influences
through dynamical considerations. In recent years, a number of
developments have enabled modeling methodology to keep pace. Aims
of research that combine information from social sciences and
biology/medicine are to setting policies to alarm, slow and
contain a possible pandemic. For highly transmissible diseases
policies may include social distancing \emph{deleted policies},
nurseries and schools closure, transport and travel restrictions,
as well as production and distribution of sufficient quantities of
vaccine. The application of those policies will have great effect
on the quality of life of the population, the job market and the
economy. An example of this coupling is given by the fact that the
lack of information from neighbors and media on the disease has
the same effect of the disease incubation period i.e. the lack of
symptoms when the virus is not demonstrable. Similarly a chronic
disease, which represent a latent but infectious state, may reduce
the level of surveillance as well as continuous media and
neighborhood alarm. Viral diseases have different intrinsic
biological characteristics which become coupled with different
social and psychological behaviors of the neighborhood, generating
a vast combinatorial of dynamics. Representative examples are
acute diseases such as the common cold (rhinovirus) and influenza
Yellow Fever; acute infection with rare late complications and
virus not readily demonstrable such as measles and SSPE; latent
persistent diseases where the virus is not readily demonstrable
such as Herpes simplex, Varicella-zoster, Measles; chronic
persistent diseases such as Hepatitis B, LCMV in Newborn Mice;
chronic infection, late diseases such as HTLV-1 leukemia and HIV;
finally, very slow infection such the prion responsible for the
Creutzfeld-Jakob disease. The probability of contacts leading to
infection can be calibrated against seasonal or environmental
effects and total and age-specific illness attack rates of data in
past pandemics. Age dependent distributions is important to take
into account whether an infected person becomes ill or remains
asymptomatic and, if symptomatic, when (if ever) the person
withdraws to household-only contacts ~\cite{Germann2006}. Glass
and collaborators~\cite{GKG2004} found that heterogeneity in
measles vaccination coverage can lead to an increased rate of
infection among non-vaccinated individuals, with a simultaneous drop
in the average age at infection. Note that since measles and
pertussis are first world vaccine-preventable diseases, the public
health problem in developing countries will affect the contact
networks in a different ways from first world. A major factor is
the correct identification of target age groups. Recent works show
that pre-scholar children aged 3 to 4 drive influenza epidemics
and are most strongly linked with mortality in the vulnerable
groups (elderly) and general population than other
children~\cite{BM2005}. In fact they present flu-like respiratory
illness as early as late September, while children aged 0-2 began
arriving a week or two later and older children first arrived in
October and adults began arriving only in November. This example
points to the difference between high-risk individuals, for
example babies under 24 months or the elderly, and those who are
transmitting the disease to everyone else . The former should be
vaccinated first~\cite{A2006}. The above examples show that the
field is at the early stage and will benefit from an
interdisciplinary approach and from a methodic and careful analysis
of the contribution of each parameter.

\end{document}